\documentclass[journal,10pt]{IEEEtran}
\usepackage{url}
\usepackage{enumerate}
\usepackage{graphicx}
\usepackage{amsmath,amssymb}
\usepackage{bm}
\usepackage[algo2e]{algorithm2e}
\usepackage{algorithm,algpseudocode}
\usepackage{multirow}
\begin{document}

\title{Towards 5G: Joint Optimization of Video Segment Cache, Transcoding and Resource Allocation for Adaptive Video Streaming in a Muti-access Edge Computing Network}

\author{Xinyu Huang, Lijun He$^{*}$, Liejun Wang, Fan Li

\thanks{X. Huang, L. He and F. Li are with the School of Information and Communications Engineering, Xi'an Jiaotong University, Xi'an 710049, China (e-mail: xinyu\_huang@stu.xjtu.edu.cn; jzb2016125@xjtu.edu.cn; lifan@mail.xjtu.edu.cn). \textit{(Corresponding author: Lijun He, email: jzb2016125@xjtu.edu.cn)}

L. Wang is with the College of Information Science and Engineering, Xinjiang University, Urumqi 830046, China (email: wljxju@xju.edu.cn).}}

{}

\maketitle

\begin{abstract}
The cache and transcoding of the multi-access edge computing (MEC) server and wireless resource allocation in eNodeB interact and determine the quality of experience (QoE) of dynamic adaptive streaming over HTTP (DASH) clients in MEC networks. However, the relationship among the three factors has not been explored, which has led to limited improvement in clients' QoE. Therefore, we propose a joint optimization framework of video segment cache and transcoding in MEC servers and resource allocation to improve the QoE of DASH clients. Based on the established framework, we develop a MEC cache management mechanism that consists of the MEC cache partition, video segment deletion, and MEC cache space transfer. Then, a joint optimization algorithm that combines video segment cache and transcoding in the MEC server and resource allocation is proposed. In the algorithm, the clients' channel state and the playback status and cooperation among MEC servers are employed to estimate the client's priority,  video segment presentation switch and continuous playback time. Considering the above four factors, we develop a utility function model of clients' QoE. Then, we formulate a mixed-integer nonlinear programming mathematical model to maximize the total utility of DASH clients, where the video segment cache and transcoding strategy and resource allocation strategy are jointly optimized. To solve this problem, we propose a low-complexity heuristic algorithm that decomposes the original problem into multiple subproblems. The simulation results show that our proposed algorithms efficiently improve client's throughput, received video quality and hit ratio of video segments while decreasing the playback rebuffering time, video segment presentation switch and system backhaul traffic.

\end{abstract}

\begin{IEEEkeywords}
MEC, QoE, video segment cache and transcoding, resource allocation, cooperation among MEC servers, Branch and Bound.
\end{IEEEkeywords}

\IEEEpeerreviewmaketitle

\section{Introduction}

The recent rapid increase in mobile data traffic and smart terminal devices has placed higher requirements on the architecture and capacity of mobile networks. For companies such as YouTube and Facebook Video, which account for more than 45$\%$ of the peak download traffic in North America \cite{Sandvine}, video streaming applications represent a significant share of internet traffic. Cisco predicts that video traffic will represent 79$\%$ of all consumer traffic by 2022 \cite{Cisco}. The delivery of video content, especially ultra high-definition video, has been considered as a critical scenario in the fifth generation (5G) network \cite{Liu_Chen}. As the key technology of the 5G mobile communication system \cite{ETSI}, the MEC server is placed close to client nodes to obtain client information (e.g., client requirements, network state and playback status) in real time. The burden of video traffic from the core network can be reduced dramatically by deploying the MEC server in the edge network. MEC-based video services provide the following advantages compared with traditional video delivery methods. 1) Precache popular videos: Based on the clients’ request and channel state, the MEC server will precache the suitable presentation of videos for clients to download and provide these videos to the nearby base stations for clients to access. The advantage is that this approach can avoid repeated video content transmission and decrease the burden on the backhaul network, thereby achieving faster service response. 2) Fast video transcoding: Due to the powerful computing capacity of the MEC server, it can transcode a video into several video sequences of different presentations and quickly respond to the client's time-varying channel state to satisfy the requirements of clients with different channel states and video requests to improve their QoE. 3) Context awareness: The MEC server can obtain the client’s channel state and playback status in real time and employ the video cache and transcoding strategy to provide a suitable resource allocation strategy for the base station, which can enhance the utilization of wireless resources to improve the client's throughput and playback continuity. In conclusion, how to utilize the advantages of the MEC server to improve the client's QoE while decreasing backhaul traffic and transmission delay is an urgent problem.

The authors in \cite{Taleb} analyzed the MEC framework and reference structure proposed by the European Telecommunications Standards Institute (ETSI). The function modules and key parameters of the MEC framework were described in detail from the perspective of video streaming service optimization. This work can be regarded as a guide for future research on MEC-based video streaming service optimization. In another respect, \cite{Yang} proposed an evaluation indicator for MEC-based video streaming service and established an accurate prediction model for video popularity and channel quality to improve the MEC-based video streaming service. In addition, the authors in \cite{Wang_Peng} presented the application of MEC in the DASH service and discussed the challenges and prospects of video streaming service optimization. Based on the above work, the key parts of MEC-based video streaming optimization can be summarized as optimization of video segment cache, transcoding and wireless resource allocation.

In general, some problems remain to be solved: (1) The video segment delete strategy ignores the consistency between the remaining segment presentation and the client's transmission capacity requirement. (2) The importance of the first few segments of a video has not been considered, which may result in a large initial delay. (3) The client’s playback status and channel state have been neglected, which leads to frequent rebuffering events. (4) Collaboration among MEC servers has not been exploited; thus, video information is not shared. (5) The resource allocation strategy is optimized independently of the MEC cache and transcoding.

Inspired by the above problems, we focused on the MEC video segment cache update strategy and the joint optimization of the MEC cache, transcoding and wireless resource allocation to improve the system throughput, video quality received by clients and video segment hit ratio while reducing the client's playback rebuffering time, presentation switch and system backhaul traffic. Our work is novel in the following aspects:

\textit{1)Comprehensive MEC-based Video Streaming Framework}

In contrast to the existing independent optimization frameworks for video segment cache, transcoding, and resource allocation optimization, we establish a comprehensive framework that includes video segment cache, transcoding in the MEC server, and resource allocation optimization in the eNodeB in the multi-MEC server collaboration scenario. The establishment of this framework makes full use of the interaction between the video segment cache and transcoding in the MEC server and resource allocation in the eNodeB to improve the client's QoE.

\textit{2)	Refined Video Segment Updated Delete Strategy}

According to the characteristics of each segment, the MEC cache is divided into a high-popularity and high-importance cache, a high-popularity but low-importance cache and a low-popularity cache to avoid additional backhaul traffic when frequent updates occur. To avoid the situation that the cached video segment does not match the client's transmission capability, we combine the number of video segment requests in the current update period with the client's transmission capacity to calculate the delete priority for the video segment delete strategy.

\textit{3)	Joint Optimization of Video Cache, Transcoding in the MEC Server and Resource Allocation in the eNodeB}

Rather than performing the video segment cache, transcoding strategy in the MEC server and resource allocation strategy in the eNodeB independently, video segment cache, transcoding and resource allocation are formulated into a mathematical model to be jointly optimized based on the clients' channel state and playback status and the cooperation among MEC servers. A utility function model, which consists of the client’s priority, the video segment switching magnitude, and the continuous playback time, is established to improve the clients' QoE. Moreover, a cost function that integrates video segment cache, transcoding and resource allocation is developed to solve the formulated problem. The relationship among the resource block (RB) assignment, the modulation and coding scheme (MCS) selection and the video segment cache and transcoding is fully utilized rather than performing only MCS selection and RB assignment with a fixed block error rate (BLER) or determining the proportion of the wireless resource that each client can access.

\section{Related Work}
To optimize the QoE of the video streaming service of DASH clients, MEC can provide services such as video segment cache and transcoding in real time. Furthermore, the eNodeB allocates wireless resources to clients based on the cached and transcoded video segments.

\begin{itemize}
 \item[$\blacktriangleright$] \textit{MEC Video Segment Cache Optimization Algorithms}
\end{itemize}

Researchers have proposed many MEC video cache optimization algorithms, which can be divided into two categories: client QoE-driven video cache optimization algorithms and cache hit ratio-driven MEC cache optimization algorithms.
 
\textit{1)	Client QoE-Driven MEC Cache Optimization Algorithms}

The authors in \cite{Li_Toni} considered the rate-distortion characteristics of different videos, video popularity, the transmission capacity of the base station and the MEC server, and the client's initial delay and cached the best video presentation for each MEC server to improve the client's QoE. To further improve the client's QoE, \cite{Mehrabi_QoE} selected the video rate, initial delay, interruption time, presentation switch and fairness as the cache update indicators and explored the trade-off between QoE and backhaul traffic. However, the fairness index neglected the average clients' rebuffering time. Considering the effect of video quality fluctuation on a client's QoE, the authors in \cite{Ge} attempted to cache the video presentation that can be supported according to the throughput of the current network and guaranteed minimization of video quality fluctuation. However, the disadvantage is that the delete policy made decisions based on only the number of segment requests in the current update period without considering the overall popularity, which may cause an additional burden on the backhaul network. To reduce the burden on the backhaul network, \cite{Tuysuz} proposed a collaborated cache among MEC servers and a QoE-aware video cache update algorithm to decrease video quality fluctuation and backhaul traffic. However, the client’s playback status was not considered in the cache optimization, which may cause rebuffering events. 
 
\textit{2)	Cache Hit Ratio-Driven MEC Cache Optimization Algorithms}

The authors in \cite{Chen_He} established a video request probability model based on video popularity, clients' preferences and video presentation characteristics to improve the video cache hit ratio, but the cooperative cache among MEC servers was neglected. In \cite{Tran_Coll}, the authors considered the cooperative cache among MEC servers. The video cache hit ratio was improved by caching multiple presentations of videos and real-time transcoding, but caching complete videos reduces the utilization of MEC cache space. In addition, the authors in \cite{Tran_Ada} combined cooperative cache with a video replacement strategy to increase the video cache hit ratio and decrease the client's initial delay. To further improve the efficiency of the MEC cache strategy, the authors in \cite{Jiang} proposed a Multi-Agent Reinforcement Learning (MARL)-based cooperative content caching policy for the MEC server to improve the video hit ratio when the users' preference and video popularity are unknown. However, the computing capacity of the MEC server was not fully exploited to transcode more video presentations that can satisfy the clients' transmission requirements. In general, the current MEC cache update optimization algorithm cannot fully exploit the information of the client's transmission state, playback status, video segment content characteristics, and cooperation cache among MEC servers. Therefore, the client's QoE in the MEC video streaming service requires further improvement.

\begin{itemize}
	\item[$\blacktriangleright$]\textit{ MEC Video Transcoding Optimization Algorithms}
\end{itemize}

Since the MEC server has powerful computing capacity, a high-presentation video segment can be quickly transcoded into a low-presentation segment to meet the client's transmission requirements. Most studies, such as \cite{Mok}\cite{Thang}, have utilized the predicted channel state and client's requirements to realize the transcoding service, thereby improving the client's QoE. The authors in \cite{Zheng} proposed several online transcoding schemes that transcoded video segments into low-presentation segments as much as possible to save the computing resources of the MEC server. To further exploit the computing capacity of the MEC server, \cite{Krish} employed a Markov prediction model to predict the rate information of the next segment requested by the client, which can enhance the pretranscoding efficiency. In addition, the authors in \cite{Li} provided a transcoding platform for video streaming service by utilizing computing resources to offer a suitable video rate in the case of ensuring the quality of service (QoS) level but ignored the effect of presentation switch on the client's QoE. To further improve the client's QoE, the authors in \cite{Wang_Peng} designed an online transcoding MEC framework by utilizing the characteristic that the MEC server is close to the clients and can obtain the client’s channel state in real time; however, cooperation among MEC servers should also be considered. Then, the authors in \cite{Wang_Sun} considered a cooperation mechanism among MEC servers and determined the video presentation to be cached based on the client's preference and historical requested video rates to reduce the consumption of computing resources of the MEC server. In addition, the authors in \cite{Liu} proposed a joint video transcoding and delivery algorithm in the case of precaching all presentations of the most popular videos and the low presentation of the low-popularity videos to decrease the client's delay. In conclusion, the current MEC-based transcoding work has not been associated with the MEC cache strategy and the resource allocation strategy in eNodeB, which results in failure to guarantee a high client QoE.

\begin{itemize}
	\item[$\blacktriangleright$]\textit{ QoE-Driven Wireless Resource Allocation Algorithms}
\end{itemize}

In addition to the MEC-based cache and transcoding strategy, wireless resource allocation is a key factor influencing client QoE. Many wireless resource allocation algorithms to improve the client's QoE in cloud-based video streaming services exist. The authors in \cite{Li_Li} narrowed the range of resource allocation by dividing clients into four types based on their buffer time to decrease the average playback rebuffering time of clients. However, the proposed greedy algorithm still cannot obtain the optimal resource allocation. Considering the client's experience and the operator's revenue, the impact of different resource allocation criteria on the fairness and video quality was analyzed in \cite{Park}, which formed a Nash balance problem to improve the video quality of more clients while ensuring the operator's revenue. In addition, the QoS of transmission also influenced the client's QoE. In \cite{He_Shan}, different channel allocation schemes for different client types (high-speed mobile clients and background clients) were proposed to allocate more channel resources to high-speed mobile clients as much as possible while ensuring the QoS requirements of background clients. \textit{Additionally, many wireless resource allocation algorithms in the MEC-based video streaming service have been proposed to fully exploit the computing capacity and cache capacity of the MEC server.} The authors in \cite{Frang} proposed an MEC video cache delivery framework to assist the base station in completing resource allocation, which can reduce the client's video playback rebuffering time. To further improve the client's QoE, \cite{Xu} proposed an MEC-enhanced ABR video delivery scheme that can flexibly adjust the video bit rate delivered to clients according to the transmission capacity, but the cooperation mechanism among MEC servers can be utilized to provide clients with the requested video segment in time. In addition, the authors in \cite{Mehrabi} proposed a QoE-driven MEC cache and resource allocation algorithm that performs load balancing on the resource blocks in the base station to avoid client access congestion. In general, the joint optimization of video segment cache and transcoding in the MEC server and resource allocation in the eNodeB is not considered in the current MEC-based video streaming optimization schemes. Therefore, we performed a preliminary exploration by selecting the client's priority, the switch range of video segment presentation, the playback time that received frames can support and the quality of received frames to form the function model and decompose the original problem into multiple subproblems. The objective of our mathematical model is to improve the client's throughput, received video quality and hit ratio of the video segment while decreasing the playback rebuffering time and system backhaul traffic.

\section{System Model}
\subsection{System Framework}
\begin{figure}[!h]
	\centering
	\includegraphics[width=8.8cm]{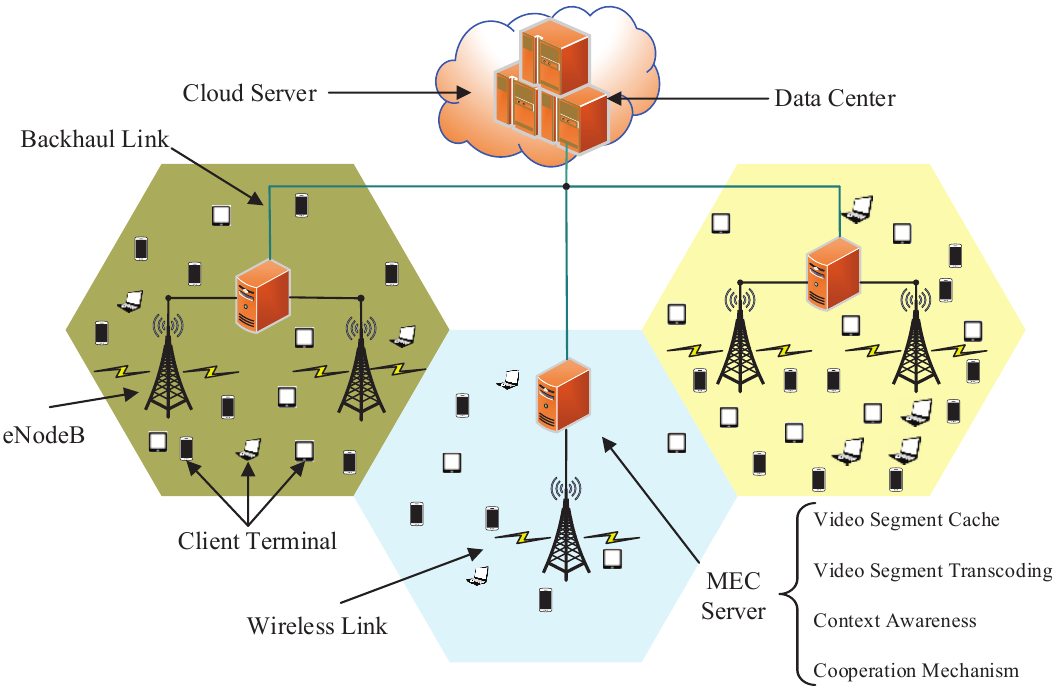}
	\caption{Video Streaming Service in the MEC Assisted Network}
	\label{fig:video_services}
\end{figure}

In this paper, we consider a scenario where multiple clients can request video streams from an eNodeB. The eNodeB can download the requested video segment from the local MEC server, the cooperative MEC servers and the cloud server, as shown in Fig. \ref{fig:video_services}. The function of the MEC server consists of video segment cache, transcoding, context awareness and a cooperation mechanism, which can utilize the information of the client's playback status and channel state to make the decision on the video segment cache and transcoding strategy and provide the wireless resource allocation strategy for eNodeB.

\begin{figure}[!h]
	\centering
	\includegraphics[width=8.8cm]{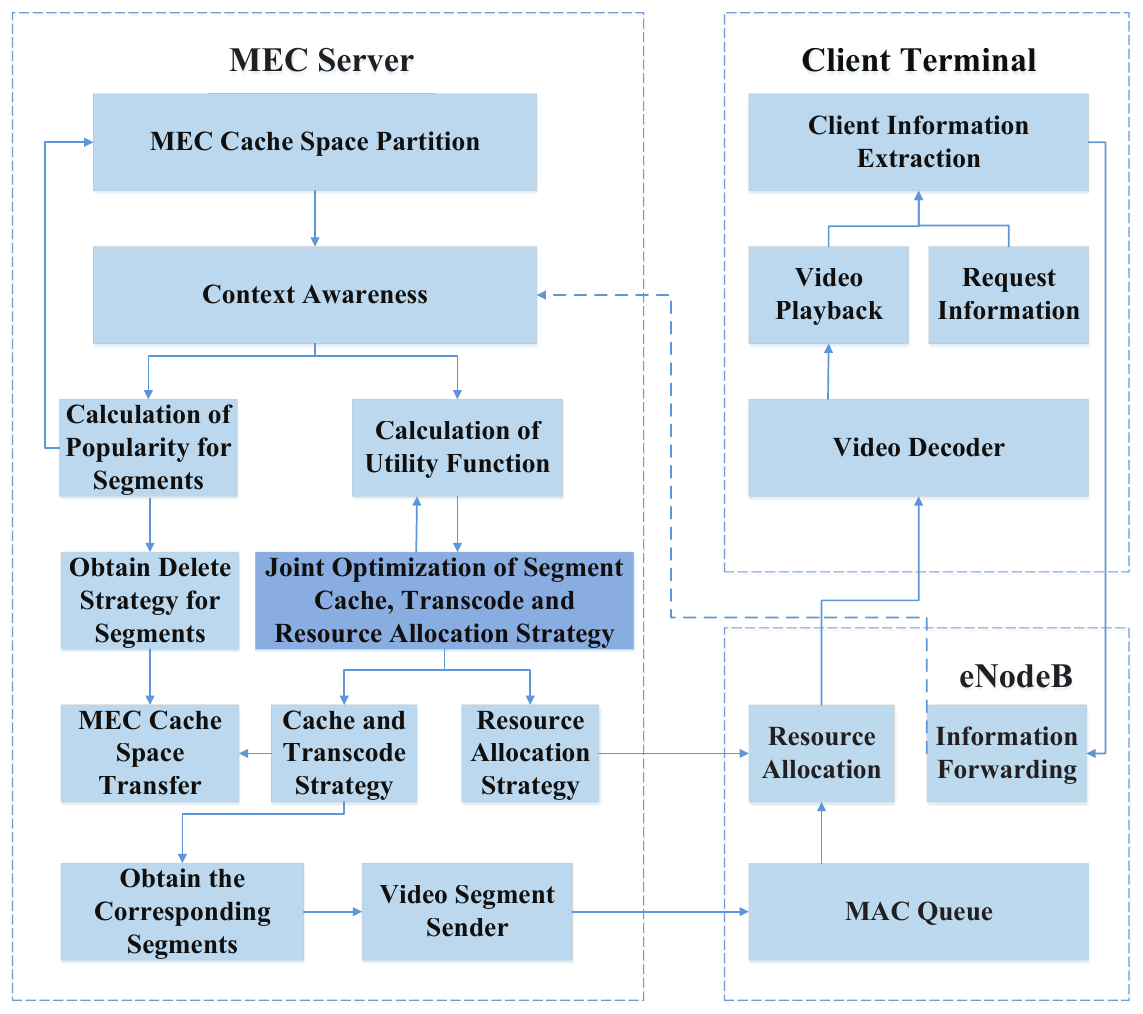}
	\caption{System Framework}
	\label{fig:system}
\end{figure}

Since the MEC server has powerful cache and computing capacity, it can quickly make decisions about the video segment cache, transcoding and resource strategy, as shown in Fig. \ref{fig:system}. First, the information of the client side, i.e., requested video, playback status, channel state, etc., is sent to the local eNodeB and then forwarded to the local MEC server in real time. The local MEC server divides the MEC cache into three different areas by calculating the video segment popularity based on the number of video segment requests, clients' playback status and channel state. In addition, the local MEC server calculates the client's utility function according to the client's playback status, channel state, and cooperation among MEC servers. The goal is to maximize the utility of all clients by jointly optimizing the video segment cache, transcoding and wireless resource allocation strategy. Then, based on the video segment cache and transcoding strategy, the local MEC server caches the corresponding video segments from the cooperative MEC servers and the cloud server or transcodes the cached video segment to one that is more suitable for transmission. The eNodeB adjusts the RBs assigned to clients according to the wireless resource allocation strategy determined by the local MEC server to improve clients' QoE.

\subsection{Refined MEC Cache Mechanism}
Since the clients continuously request new video segments and the cache space of the MEC server is limited, the clients may suffer rebuffering events if the requested video segment cannot be cached in time. Therefore, in each cache update period, the client's request and channel state are utilized to update the video segment popularity, and low-popularity video segments are deleted sequentially to guarantee sufficient space to cache new segments for clients.

In this paper, we select the video segment as the smallest unit in the MEC cache mechanism. In each cache update period, the delete priority of each video segment is calculated as the basis for the video segment delete strategy. The MEC cache mechanism consists of MEC cache predeployment, MEC cache partitioning, video segment delete strategy, MEC cache space transfer and a cooperation mechanism.

\subsubsection{Historical Requested Video Data-Based MEC Cache Predeployment}
When deploying the MEC servers, the size of the cache space in each area must be determined. More cache resources should be allocated to areas where the amount of historical requested video data is high to avoid the insufficiency of cache capacity.

We denote the set of MEC servers by $\mathbb{Q}=\{1,2,...,Q\},q\in \mathbb{Q}$. Each MEC-served region ${{\Re }_{q}}$ with ${{K}_{q}}$clients is served by one MEC server and several eNodeBs. The set of eNodeBs served by the MEC server $q$ is ${{\mathcal{H}}_{q}}=\{1,2,...,{{H}_{q}}\}, {{h}_{q}}\in {{\mathcal{H}}_{q}}$, and the client set of eNodeB ${{h}_{q}}$ is $\mathbb{Z}({{h}_{q}})=\{1,2,...,{{K}_{{{h}_{q}}}}\}$,$k\in \mathbb{Z}({{h}_{q}})$. The requested video data in one month are generated based on the video request frequency of different clients. Let us denote the set of video files by $\mathcal{F},f\in \mathcal{F}$ and the total cache space of MEC servers by $S$. Then, we can analyze the requested video data in one month to allocate the cache space to each MEC server, as follows:

\begin{equation}
{{S}_{q}}=\frac{\sum\limits_{k\in {{\Re }_{q}}}{\sum\nolimits_{f\in \mathcal{F}}^{{}}{\sum\nolimits_{l=1}^{L}{\sum\nolimits_{i=1}^{S{{R}_{k,f}}}{\Gamma _{k,f,i}^{l}s_{f,i}^{l}}}}}}{\sum\limits_{k\in \Re }{\sum\nolimits_{f\in \mathcal{F}}^{{}}{\sum\nolimits_{l=1}^{L}{\sum\nolimits_{i=1}^{S{{R}_{k,f}}}{\Gamma _{k,f,i}^{l}s_{f,i}^{l}}}}}}S
\end{equation}
where ${{S}_{q}}$ is the size of the allocated cache space for MEC server q and $\Gamma _{k,f,i}^{l}$ is the number of times that client $k$ requests presentation $l$ of video segment $i$ of video $f$. $s_{f,i}^{l}$ and $S{{R}_{k,f}}$ denote the file size of presentation l of video segment $i$ of video $f$ and the maximum index of video segment $f$ requested by client $k$, respectively.

\subsubsection{MEC Cache Partitioning Based on Video Content Characteristics and Popularity}
The statistical results show that the cumulative probability of a client watching the first 15\% of the video content and then leaving is 0.5, which indicates that the first 15\% of the video content is most important \cite{Chen}. We denote the number of the first 15\% of video segments of video $f$ by $e_f$ and the number of the whole video segments of video $f$ by${{N}_{f}}$, where the relationship between ${{e}_{f}}$ and ${{N}_{f}}$ is ${{e}_{f}}=\left\lceil 0.15{{N}_{f}} \right\rceil $.

If all stored video segments are updated in real time, the video segments deleted in the current period may need to be cached in the next period, which increases the backhaul traffic. Therefore, we divide the MEC cache into three categories and employ ${{\Delta }_{1}},{{\Delta }_{2}},{{\Delta }_{3}}$ to indicate that (1) the first 15\% of video segments of the 20\% most popular videos constitute the first part of the cache ${{\Delta }_{1}}$ (The 20\% most popular video accounts for 80\% of web traffic, which can be updated less to alleviate backhaul traffic \cite{Cha}), (2) the remaining video segments of the 20\% most popular videos constitute ${{\Delta }_{2}}$, and (3) the video segments of the last 80\% most popular videos compose  ${{\Delta }_{3}}$. Cache ${{\Delta }_{1}}$ is updated in every long period $\mathcal{J}$($\mathcal{J}\in \{1,2,...,\Theta \}$), and caches ${{\Delta }_{2}}$ and ${{\Delta }_{3}}$ are updated in every short period $\gamma$ ($\gamma\in\{1,2,...,\Upsilon\}$), as shown in Fig. 3.

\begin{figure}[!h]
	\centering
	\includegraphics[width=9cm]{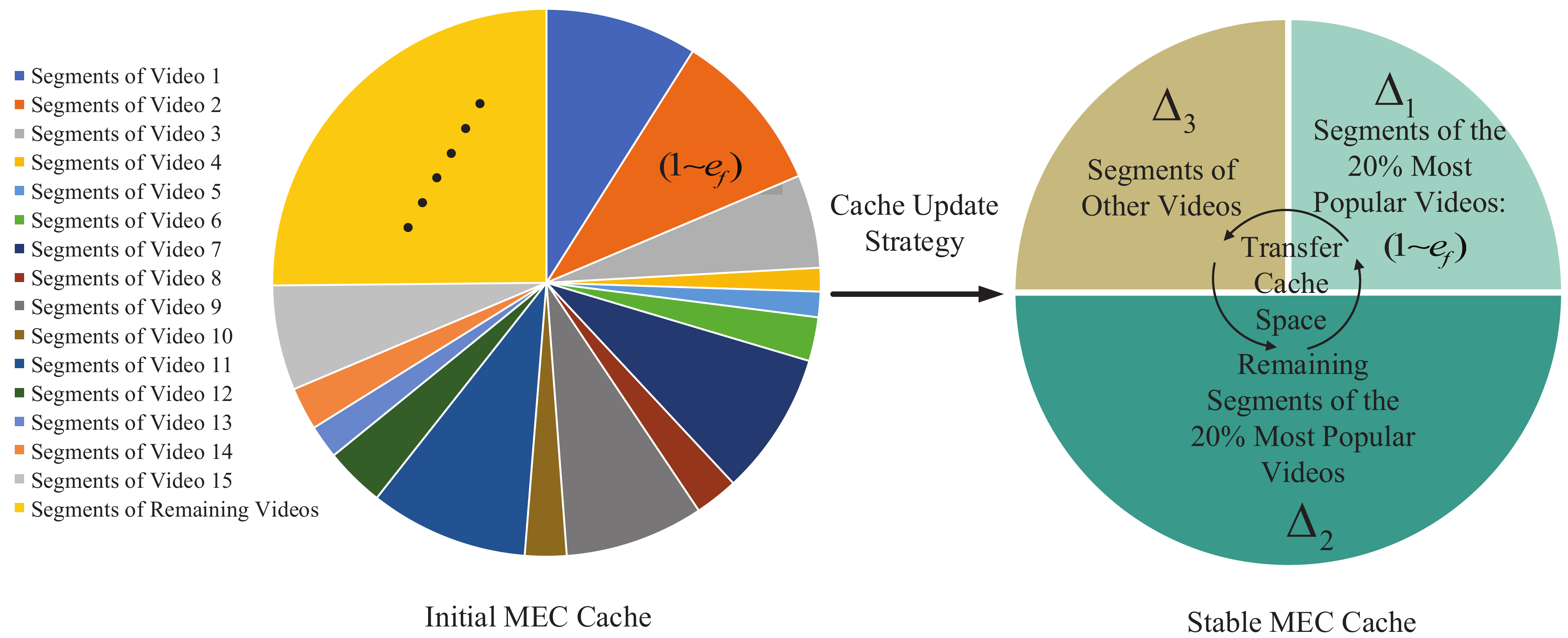}
	\caption{MEC Cache Partitioning Process}
	\label{fig:Cache}
\end{figure}

There are two processes in Fig. \ref{fig:Cache}, the initial MEC cache and the stable MEC cache after the cache update strategy. In the initial time, since the number of requested videos is large and the MEC cache space is limited, the first $e_{f}$ video segments of the highest popularity presentation of video f are placed in the MEC cache space sequentially based on descending video popularity until the MEC cache space is full to decrease the clients' initial delay and improve their QoE. Since the MEC will perform a cache update strategy in every update period, the MEC cache can be divided into three parts according to the type of video segment, and the size of each subcache will dynamically change based on the stored video segments. Let us denote the size of cache ${{\Delta }_{1}}$ by $S{{H}_{q}}$ and the total size of cache ${{\Delta }_{2}}$ and ${{\Delta }_{3}}$ by $S{{C}_{q}}$.

\subsubsection{Video Segment Delete Mechanism Based on Client Request and Channel State}
During each cache update period, only those segments with high delete priority are deleted, instead of deleting the entire video, so that the MEC sever can retain more video segments. When the client is stuck or about to be stuck, the MEC server can quickly provide the needed segment to improve the client's QoE.

First, we number the set of video segments in each MEC server. Then, the requested number, client transmission capacity and segment rate are employed to obtain the delete priority of the different presentations of the video segments in period $\gamma$ for MEC server $q$, which can be depicted as
\begin{equation}
DP_{i}^{\gamma ,q,l}=\frac{1}{{{K}_{q}}}\sum\nolimits_{k=1}^{{{K}_{q}}}{\frac{1}{g_{k,i}^{\gamma ,l}+\zeta \cdot \upsilon _{k,i}^{l}+\alpha }}
\end{equation}
where $\zeta$ and $\alpha$ are positive variables and $g_{k,i}^{\gamma ,l}$ indicates whether client $k$ requests presentation $l$ of video segment $i$ in period $\gamma$. If it is requested, $g_{k,i}^{\gamma ,l}=1$; otherwise, $g_{k,i}^{\gamma ,l}=0$. $v_{k,i}^{l}$ reflects the match relationship between the average transmission capacity, $\widehat{CP}_{k,i}^{l}$, and the video segment rate, $R_{i}^{l}$. 

\begin{equation}
 \upsilon _{k,i}^{l}=
 \left\{ \begin{aligned}
& 1,\widehat{CP}_{k,i}^{l}\ge R_{i}^{l} \\ 
& 0,otherwise \\ 
\end{aligned}\right.          
\end{equation}

However, a delete strategy based on only the delete priority in the current period $\gamma$ may result in the deletion of video segments with low delete priority in the previous periods. Therefore, we employ the delete priority of the previous period to update the current period as follows:

\begin{equation}
\widehat{DP}_{i}^{\gamma ,q,l}=\lambda \widehat{DP}_{i}^{\gamma ,q,l}+(1-\lambda )\widehat{DP}_{i}^{\gamma -1,q,l}
\end{equation}
where $\widehat{DP}_{i}^{\gamma ,q,l}$ is the overall delete priority of video segment $i$ of presentation $l$ in MEC server $q$ up to period $\gamma$. $\lambda $ is a positive constant between 0 and 1. 

The video segment update strategy varies in different caches. For cache $\Delta_{1}$, since the first 15\% of video segments of video f are more important to clients, we update them according to the latest video segment popularity in each period $\mathcal{J}$. For caches $\Delta_{2}$ and $\Delta_{3}$, in each update period $\gamma$, the stored video segments are sequentially deleted based on the overall delete priority of video segments, excluding the video segments being transmitted until the size of the cache is sufficient to store the video segments to be downloaded.

\subsubsection{MEC Cache Space Transfer}
The segments in $\Delta_1$ are updated in every period $\mathcal{J}$. Due to the length and code rate, the size of video segments differs greatly, which results in the size of the deleted video segments not being equal to that of the downloaded video segments. To make full use of each cache space, we compare the size of $\Delta_{2}$ before and after each update period to release or increase space in $\Delta_2$ and $\Delta_3$. In the beginning of each period $\mathcal{J}$, the change in the size of $\Delta_1$ is derived by

\begin{equation}
FC_{q}^{\mathcal{J}}=\sum\limits_{f\in \mathcal{F}_{pop}^{\mathcal{J}}}{\sum\limits_{i=1}^{{{e}_{f}}}{\sum\limits_{l=1}^{L}{s_{f,i}^{l}}}}-SH_{q}^{\mathcal{J}-1} 
\end{equation}
where $\mathcal{F}_{pop}^{\mathcal{J}}$ denotes the set of the 20\% most popular videos in period $\mathcal{J}$ and $SH_{q}^{\mathcal{J}\text{-1}}$ denotes the size of $\Delta_1$ in period $\mathcal{J}-1$. Then, the sum of the sizes of $\Delta_2$ and $\Delta_3$ can be updated according to 
\begin{equation}
SC_{q}^{\mathcal{J}}=SC_{q}^{\mathcal{J}-1}-FC_{q}^{\mathcal{J}}
\end{equation}

When $FC_{q}^{\mathcal{J}}\ge 0$, the video segments in $\Delta_{2}$ and $\Delta_{3}$ are deleted according to the delete priority to release the cache to $\Delta_{1}$ until the size of the cache of $\Delta_{1}$ is satisfied.

\subsubsection{Cooperation Mechanism among MEC Servers}
Provided that the neighboring MEC servers store the requested segments, the local MEC server will download them from the neighboring MEC server $p$ with the highest remaining transmission capacity. The sum of the size of downloaded segments should not exceed the transmission capacity between the local MEC server q and the neighboring MEC server $p$.
\begin{equation}
\sum\limits_{k=1}^{{{K}_{q}}}{\sum\limits_{l=1}^{L}{\tau _{k,i,l}^{q}\chi _{k,i,l}^{p}s_{k,i}^{l}}}\le CP_{q,p}\cdot TD,\forall q\in \mathbb{Q},p\in \mathbb{Q}\backslash \{q\}               
\end{equation}
where $\tau _{k,i,l}^{q}$ denotes the optimized variable, indicating whether the requested video segment will be cached in MEC server $q$. If it will be cached in MEC server $q$, $\tau _{k,i,l}^{q}=1$; otherwise, $\tau _{k,i,l}^{q}=0$. $\chi _{k,i,l}^{p}$ denotes whether the requested video segment is stored in MEC server p. If it is stored, $\chi _{k,i,l}^{p}=1$; otherwise, $\chi _{k,i,l}^{p}=0$. $CP_{q,p}$ is the transmission capacity between MEC servers $q$ and $p$. $TD$ is the time duration of period $\gamma$. If the shared video segments exceed the transmission capacity, the video segment sharing mechanism among MEC servers will be stopped, and the local MEC server $q$ must request the video segment from the cloud server or wait until the next update period.

\section{Utility Function Model Based on Client’s QoE}
For the MEC server, the client's playback status and channel state are particularly important for determining the cache, transcoding and resource allocation strategies. Providing the client with the appropriate presentation of the video segment can improve the client's QoE, and precaching video segments required by most clients can reduce the traffic burden on the backhaul network. Therefore, establishment of the client's QoE-based utility function model is critical for the cache, transcoding and resource allocation strategies. Next, we formulate the function model based on the client's priority, the switching magnitude of video segment presentation, the playback time that received frames can support and the quality of received frames.
\subsection{Factors of the Utility Function}
\subsubsection{Client's Priority}
To better distinguish each client's urgency according to the corresponding playback buffer status, a segmentation exponential function is used to indicate the urgency of client k requesting video segment i, as follows:
\begin{equation}
EL_{k,i}^{\gamma }=exp(\frac{\omega }{1+BT_{k,i-1}^{\gamma }/{{d}_{k,i-1}}})-1
\end{equation}
where $exp(\bullet)$ is the exponential function. $BT_{k,i-1}^{\gamma }$ is the remaining buffer time when client $k$ receives video segment $i-1$ of presentation $l$ at the end of period $\gamma$. ${{d}_{k,i-1}}$ is the time duration of video segment $i-1$ requested by client $k$, and $\omega $ is a positive constant. Since the times required to acquire video segments from the cloud server and the neighboring MEC server are different, when the requested video segment is downloaded from the cloud server, the urgency of client $k$ increases because of the increased download time. Therefore, we define the client's priority based on the client's urgency as follows:
\begin{equation}\small
Pr(\tau _{k,i,l}^{q},o_{k,i,l}^{q})=(\tau _{k,i,l}^{q}+o_{k,i,l}^{q})(EL_{k,i}^{\gamma }+EL_{k,i}^{\gamma }\prod\limits_{p}{(1-\chi _{k,i,l}^{p})})
\end{equation}
 
\subsubsection{Presentation Switching Magnitude} 
During the client's playback, frequent presentation switches of video segments will seriously affect the client's QoE. Therefore, without exceeding the cache capacity and transcoding capacity of the MEC server, the presentation of requested video segment $i$ of client $k$ should remain consistent with video segment $i-1$ as much as possible by cache or transcoding. The function of the presentation switching magnitude can be depicted as

\begin{equation}
RS\left( \tau _{k,i,l}^{q},o_{k,i,l}^{q} \right)=\frac{(\tau _{k,i,l}^{q}+o_{k,i,l}^{q})|l-{{l}^{'}}|}{{{l}_{\max }}-{{l}_{\min }}}
\end{equation}
where $\left| \bullet  \right|$ denotes the absolute function and ${{l}^{'}}$ is the presentation of video segment $i-1$ of client $k$. The value range of function $RS(\bullet)$ is [0,1]. $o_{k,i,l}^{q}$ is the transcoding variable, indicating whether the requested video segment will be obtained by transcoding in MEC server $q$. If it is obtained by transcoding, $o_{k,i,l}^{q}=1$; otherwise, $o_{k,i,l}^{q}=0$. $p_{k,i,l}^{{}}$ is the transcoding capacity for client k to obtain the requested video segment, which can be depicted as $p_{k,i,l}^{{}}=\mu \cdot {{s}_{k,i,l}}$ \cite{Tran_Coll}. $\mu $ is a positive value. In each update period $\gamma$, the transcoding capacity of the MEC server offered to the clients cannot exceed the computing resources of the MEC server, which can be expressed as

\begin{equation}
\sum\limits_{k\in {{\Re }_{q}}}{\sum\limits_{l}{o_{k,i,l}^{q}p_{k,i,l}^{{}}}}\le {{C}_{q}}\times TD                          
\end{equation}
where ${{C}_{q}}$ is the computing capacity per time unit of MEC server $q$.

\subsubsection{Continuous Playback Time That Received Frames Can Support}
The greater the number of video frames received by the client is, the longer the continuous playback time that can be supported. In the update period $\gamma$, the client's transmission capability can be determined by the RB allocation to the client, which can be expressed as

\begin{equation}
c_{k}^{{{h}_{q}}}=\sum\limits_{n\in \Omega ({{h}_{q}})}^{{}}{a_{k,n}^{{{h}_{q}}}\sum\limits_{m=1}^{q(\Omega )}{b_{k,m}^{{{h}_{q}}}{{r}_{m}}}}              
\end{equation}
where $a_{k,n}^{{{h}_{q}}}$ denotes whether RB $n$ of eNodeB ${{h}_{q}}$ is allocated to client $k$. If it is allocated, $a_{k,n}^{{{h}_{q}}}=1$; otherwise, $a_{k,n}^{{{h}_{q}}}=0$. $b_{k,m}^{{{h}_{q}}}$ denotes whether client k uses MCS m. If it is used, $b_{k,m}^{{{h}_{q}}}=1$; otherwise, $b_{k,m}^{{{h}_{q}}}=0$. 

In each update period, the received video packets of the client can be expressed as
\begin{equation}
PS_{k,i,l}^{{{h}_{q}}}=\left\{ \begin{aligned}
& c_{k}^{{{h}_{q}}}\times TD,\phi _{k,i,l}^{{}}+c_{k}^{{{h}_{q}}}\times TD\le s_{k,i,l} \\ 
& s_{k,i,l}^{{}}-\phi _{k,i,l}^{{}},otherwise \\ 
\end{aligned} \right.
\end{equation} 

where ${{\phi }_{k,i,l}}$ denotes the total size of the video packets of presentation $l$ of segment $i$ already transmitted to client $k$. Then, the ratio of playback time that the received video packets can support to the time duration of segment $i$ can be depicted as

\begin{equation}
\begin{split}
SP(\tau _{k,i,l}^{q},o_{k,i,l}^{q},c_{k}^{{{h}_{q}}})&=[ST(\phi _{k,i,l}^{{}}+(\tau _{k,i,l}^{q}\text{+}o_{k,i,l}^{q})PS_{k,i,l}^{{{h}_{q}}})\\&-ST(\phi _{k,i,l}^{{}})]/{{d}_{k,i}}
\end{split}
\end{equation}
where $ST(\bullet )$ denotes the continuous playback time interval that video packets can support in \cite{He}.

\subsubsection{Quality of Received Video Frames}
Peak signal-to-noise ratio (PSNR), which is most commonly used to measure the reconstruction quality of lossy compression codecs, is an important factor that reflects the quality of video frames and can influence client's QoE. When the client receives new video frames, the PSNR of the total received video frames can be updated as

\begin{equation}\small
\begin{split}
PSNR(\tau _{k,i,l}^{q},o_{k,i,l}^{q},c_{k}^{{{h}_{q}}})=(\tau _{k,i,l}^{q}+o_{k,i,l}^{q})\sum\limits_{z=1}^{Z(PS_{k,i,l}^{h_q})}{10\lg \frac{{{255}^{2}}}{MSE_{i,l}^{z}}}
\end{split}
\end{equation}
where $Z(PS_{k,i,l}^{{{h}_{q}}})$denotes the total received video frames when client $k$ receives the new video packets of size $PS_{k,i,l}^{{{h}_{q}}}$. $MSE_{i,l}^{z}$ is the mean square error between frame $z$ of video segment $i$ of presentation $l$ and that of the original video segment, which can be defined as

\begin{equation}
MSE_{i,l}^{z}=\frac{\sum\limits_{0\le mz\le MF_{i,l}^{z}}{\sum\limits_{0\le nz\le NF_{i,l}^{z}}{{{(Pi{{x}_{mz,nz}}-Pix_{mz,nz}^{'})}^{2}}}}}{MF_{i,l}^{z}\times NF_{i,l}^{z}}
\end{equation}
where $MF_{i,l}^{z}$ and $NF_{i,l}^{z}$ denote the height and weight of frame $z$ of video segment $i$, respectively. $Pi{{x}_{mz,nz}}$ and $Pix_{mz,nz}^{'}$ are the values of the pixel with coordinates $(mz, nz)$ in the current frame and its reference frame, respectively.
  
\subsection{Utility Function Model Based on Client's QoE}    
In each update period, we establish the utility function model of client $k$ by combining the client's priority, presentation switching range, playback time and quality of received frames, which can be determined by

\begin{equation}\small
\begin{aligned}
& U'(\tau _{k,i,l}^{q},o_{k,i,l}^{q},c_{k}^{{{h}_{q}}})={{\theta }_{1}}Pr(\tau _{k,i,l}^{q},o_{k,i,l}^{q})\text{+}{{\theta }_{2}}SP(\tau _{k,i,l}^{q},o_{k,i,l}^{q},c_{k}^{{{h}_{q}}}) \\ 
& -{{\theta }_{3}}RS(\tau _{k,i,l}^{q},o_{k,i,l}^{q})+{{\theta }_{4}}PSNR(\tau _{k,i,l}^{q},o_{k,i,l}^{q},c_{k}^{{{h}_{q}}}) \\ 
\end{aligned}
\end{equation}
where ${{\theta }_{1}},{{\theta }_{2}},{{\theta }_{3}},{{\theta }_{4}}$ are positive values. According to the relationship among $a_{k,n}^{{{h}_{q}}}$, $b_{k,m}^{{{h}_{q}}}$and $c_{k}^{{{h}_{q}}}$ described in equation (12), the utility function model of client $k$ can be expressed as $U(\tau _{k,i,l}^{q},o_{k,i,l}^{q},a_{k,n}^{{{h}_{q}}},b_{k,m}^{{{h}_{q}}})=U'(\tau _{k,i,l}^{q},o_{k,i,l}^{q},c_{k}^{{{h}_{q}}})$.

\section{Joint Optimization of Segment Cache, Transcoding and Resource Allocation}
The cache and transcoding strategy in the MEC server and the resource allocation strategy in the eNodeB together influence the client's QoE. However, the classic resource allocation algorithms cannot adjust the resource allocation strategy in a timely manner based on the remaining cache and transcoding resources of the MEC server and the client's playback status, which results in failure of the client's playback continuity to be guaranteed, even if the requested video segment has been cached or transcoded in the MEC server. To guarantee the client's playback continuity and improve the client's QoE, we jointly optimize the cache and transcoding strategy in the MEC server and the resource allocation strategy in the eNodeB.

\subsection{Problem Formulation}
In this section, we present the joint optimization of the cache and transcoding strategy in the MEC server and the resource allocation strategy in the eNodeB. In cache $\Delta_{1}$, the update strategy is that the first 15\% of video segments of the first 20\% most popular videos are updated in every period $\mathcal{J}$, which is relatively fixed. In caches $\Delta_{2}$ and $\Delta_{3}$, the mathematical model with the objective of maximizing the overall utility level of the clients, subject to the constraints of MEC storage capacity, transcoding capacity, transmission bandwidth among MEC servers and transmission capacity of the eNodeB, is as follows:
\begin{equation}
\begin{split}
\begin{aligned}
& \underset{\bm{\tau ,o,a,b}}{\mathop{\max }}\,\sum\limits_{q}{\sum\limits_{k}{\sum\limits_{l}{U(\tau _{k,i,l}^{q},o_{k,i,l}^{q},a_{k,n}^{{{h}_{q}}},b_{k,m}^{{{h}_{q}}})}}} \\ 
& \text{s}\text{.t}\text{.} \\ 
& (c1)\sum\limits_{k\in {{\Re }_{q}}}{\sum\limits_{l=1}^{L}{\tau _{k,i,l}^{q}s_{k,i,l}^{{}}\le S{{C}_{q}}}},\tau _{k,i,l}^{q}\in \{0,1\},\forall q\in \mathbb{Q} \\ 
& (c2)\tau _{k,i,l}^{q}+\chi _{k,i,l}^{q}\le 1,\chi _{k,i,l}^{q}\in \{0,1\},\forall q\in \mathbb{Q} \\ 
& (c3)\sum\limits_{k\in {{\Re }_{q}}}^{{}}{\sum\limits_{l=1}^{L}{\tau _{k,i,l}^{q}\chi _{k,i,l}^{p}s_{k,i,l}^{{}}}}\le C{{P}_{q,p}}\times TD,\\&\quad\quad\chi _{k,i,l}^{p}\in \{0,1\},\forall q\in \mathbb{Q},p\in \mathcal{Q}\backslash \{q\} \\ 
& (c4)\sum\limits_{k\in {{\Re }_{q}}}^{{}}{\sum\limits_{l=1}^{L}{\tau _{k,i,l}^{q}\prod\limits_{p}{(1-\chi _{k,i,l}^{p})}s_{k,i,l}^{{}}}}\le C{{P}_{q,cloud}}\times TD,\\&\quad\quad\forall q\in \mathbb{Q},\text{ }p\in \mathbb{Q}\backslash \{q\} \\ 
& (c5)\tau _{k,i,l}^{q}+o_{k,i,l}^{q}\le 1,o_{k,i,l}^{q}\in \{0,1\},\forall q\in \mathbb{Q} \\ 
& (c6)\sum\limits_{k\in {{\Re }_{q}}}{\sum\limits_{l=1}^{L}{o_{k,i,l}^{q}p_{k,i,l}^{q}\le {{C}_{q}}}}\times TD,\forall q\in \mathbb{Q} \\ 
& (c7)\sum\limits_{k\in \mathbb{Z}({{h}_{q}})}{a_{k,n}^{{{h}_{q}}}=1},a_{k,n}^{{{h}_{q}}}\in \{0,1\},\forall n\in \Omega ({{h}_{q}}),\\&\quad\quad{{h}_{q}}\in {{\mathcal{H}}_{q}},q\in \mathbb{Q} \\ 
& (c8)\sum\limits_{m=1}^{q(\Omega ({{h}_{q}}))}{b_{k,m}^{{{h}_{q}}}=1,b_{k,m}^{{{h}_{q}}}\in \{0,1\},\forall k\in \mathbb{Z}({{h}_{q}})},\\&\quad\quad {{h}_{q}}\in {{\mathcal{H}}_{q}},q\in \mathbb{Q} \\ 
& (c9)\sum\limits_{n\in \Omega ({{h}_{q}})}{a_{k,n}^{{{h}_{q}}}\sum\limits_{m=1}^{q(\Omega ({{h}_{q}}))}{b_{k,m}^{{{h}_{q}}}{{r}_{m}}}}\le L{{S}_{k}},\forall k\in \mathbb{Z}({{h}_{q}}), \\&\quad\quad{{h}_{q}}\in {{\mathcal{H}}_{q}},q\in \mathbb{Q} \\ 
\end{aligned}
\end{split}
\end{equation}

For the constraints, constraint (c1) implies that the video segments to be cached should not be greater than the size of cache $\Delta_{2}$ and $\Delta_{3}$. Constraint (c2) guarantees that the video segments cannot be cached repeatedly. Constraint (c3) and constraint (c4) indicate that the size of video segments to be downloaded from the neighboring MEC server and the cloud server should not exceed the corresponding transmission capacity respectively. Constraint (c5) guarantees that the video segment cannot be obtained by cache and transcoding simultaneously. Constraint (c6) indicates that the transcoding capacity for video segments cannot exceed the computing resources of the MEC server. Constraint (c7) implies that each RB must be assigned to one and only one client at each scheduling period. Constraint (c8) ensures that all the RBs assigned to the same client should use the same MCS. Constraint (c9) indicates that the transmission capacity of the assigned RBs of client $k$ should be not greater than the total size of all the packets in client $k$'s MAC queue.

\subsection{Solution: Joint Optimization of Segment Cache, Transcoding and Resource Allocation}
Problem (18) can be treated as a high-dimensional nonlinear mixed-integer optimization problem that is difficult to solve. Therefore, we decompose the problem into multiple subproblems. To maximize the overall utility of all the clients, we need to assign the limited wireless resources to the clients who are actually in need to achieve the greatest utility improvement. Problem (18) shows that the video segment cache strategy, transcoding strategy and resource allocation strategy interact with each other: problem (18) can only be solved by considering the relationship among the three factors. Consequently, we first develop a cost function of RB assignment that depends on the optimal MCS selection, video segment cache and transcoding policy. The process to obtain the optimal MCS selection, video segment cache and transcoding policy with fixed RB assignment is provided. Once the value of the cost function is calculated, the RB assignment of one RB can be determined.

\subsubsection{Cost Function of RB Assignment}	
Let $N_{k}^{{{h}_{q}}}$ denote the set of the RBs of eNodeB ${{h}_{q}}$ already assigned to client $k$. Then, the cost function $CI_{k,n}^{{{h}_{q}}}$ can be defined as the utility improvement resulting from assigning another RB $n$ to client $k$. $CI_{k,n}^{{{h}_{q}}}$ can be computed by
\begin{equation}
\begin{split}
CI_{k,n}^{{{h}_{q}}}&=\underset{{{n}^{'}}\in N_{k}^{{{h}_{q}}}\cup \{n\}}{\mathop{\max }}\,\sum\limits_{l}{U(\tau _{k,i,l}^{q},o_{k,i,l}^{q},a_{k,n}^{{{h}_{q}}},b_{k,m}^{{{h}_{q}}})}\\&-\underset{{{n}^{'}}\in N_{k}^{{{h}_{q}}}}{\mathop{\max }}\,\sum\limits_{l}{U(\tau _{k,i,l}^{q},o_{k,i,l}^{q},a_{k,n}^{{{h}_{q}}},b_{k,m}^{{{h}_{q}}})}
\end{split}
\end{equation}

As shown in equation (19), the calculation of $CI_{k,n}^{{{h}_{q}}}$ depends on the optimal determination of $(\bm{\tau} _{k}^{*},\bm{o}_{k}^{*},\bm{b}_{k}^{*})$ with RB assignment set $N_{k}^{{{h}_{q}}}$ and $N_{k}^{{{h}_{q}}}\bigcup \{n\}$. As long as $CI_{k,n}^{{{h}_{q}}}$ is obtained, the optimal determination of $a_{k,n}^{{{h}_{q}}}$ can be achieved. Next, we present the process for deriving the optimal $(\bm{\tau} _{k}^{*},\bm{o}_{k}^{*},\bm{a}_{k}^{*},\bm{b}_{k}^{*})$.

\subsubsection{Optimal $(\bm{\tau} _{k}^{*},\bm{o}_{k}^{*},\bm{b}_{k}^{*})$ Determination with Fixed RB Assignment Set $N_{k}^{{{h}_{q}}}$}
In general, more bits allocated to the client will result in better utility improvement because the number or quality of received video frames increases. Thus, we can assume that utility is an increasing function of transmission capacity. To solve the problem in equation (18), we first need to determine the optimal MCS index $b_{k,m}^{{{h}_{q}}}$ to maximize the transmission capacity that the RBs in $N_{k}^{{{h}_{q}}}$ can support and then select the optimal cache and transcoding policy to maximize the utility.

i) \textit{Determination of the optimal MCS index $\bm{b}_{k}^{*}$ with fixed RB assignment set $N_{k}^{{{h}_{q}}}$}

To maximize the transmission capacity that the RBs in $N_{k}^{{{h}_{q}}}$ can support, we need to obtain the optimal solution to the following problem:
\begin{equation}
\begin{aligned}
& \underset{{{a}_{k}},{{b}_{k}}}{\mathop{\max }}\,\sum\limits_{n\in N_{k}^{{{h}_{q}}}}{a_{k,n}^{{{h}_{q}}}}\sum\limits_{m=1}^{q(N_{k}^{{{h}_{q}}})}{(1-{{E}_{k,m}})b_{k,m}^{{{h}_{q}}}}{{r}_{m}} \\ 
& \text{s}\text{.t}\text{.} \\ 
& \sum\limits_{m=1}^{q(N_{k}^{{{h}_{q}}})}{b_{k,m}^{{{h}_{q}}}}=1,b_{k,m}^{{{h}_{q}}}\in \{0,1\} \\ 
\end{aligned}
\end{equation}  
                
As shown in equation (20), ${{E}_{k,m}}$ is the BLER when client $k$ selects MCS $m$. The solution method of $\bm{b}_{k}^{*}$ is introduced in \cite{He}.

\textit{ii) Determination of the optimal cache and transcoding $(\bm{\tau} _{k}^{*},\bm{o}_{k}^{*})$ with the fixed resource allocation strategy $({\bm{a}_{k}},\bm{b}_{k}^{*})$}

When the resource allocation strategy $({\bm{a}_{k}},\bm{b}_{k}^{*})$ is determined, the client's transmission capacity in each update period can be obtained. Then, we can obtain the optimal cache and transcoding policy $(\bm{\tau} _{k}^{*},\bm{o}_{k}^{*})$ by solving the following problem:
\begin{equation}
\begin{aligned}
& \underset{\tau ,o}{\mathop{\max }}\,\sum\limits_{q}{\sum\limits_{k}{\sum\limits_{l}{U(\tau _{k,i,l}^{q},o_{k,i,l}^{q},{\bm{a}_{k}}, \bm{b}_{k}^{*})}}} \\ 
& \text{s}\text{.t}\text{.} \\ 
& (c1)\sum\limits_{k\in {{\Re }_{q}}}{\sum\limits_{l=1}^{L}{\tau _{k,i,l}^{q}s_{k,i,l}^{{}}\le S{{C}_{q}}}},\tau _{k,i,l}^{q}\in \{0,1\},\forall q\in \mathbb{Q} \\ 
& (c2)\tau _{k,i,l}^{q}+\chi _{k,i,l}^{q}\le 1,\chi _{k,i,l}^{q}\in \{0,1\},\forall q\in \mathbb{Q} \\ 
& (c3)\sum\limits_{k\in {{\Re }_{q}}}^{{}}{\sum\limits_{l=1}^{L}{\tau _{k,i,l}^{q}\chi _{k,i,l}^{p}s_{k,i,l}^{{}}}}\le C{{P}_{q,p}}\times TD,\chi _{k,i,l}^{p}\in \{0,1\},\\&\quad\quad\forall q\in \mathbb{Q},p\in \mathbb{Q}\backslash \{q\} \\ 
& (c4)\sum\limits_{k\in {{\Re }_{q}}}^{{}}{\sum\limits_{l=1}^{L}{\tau _{k,i,l}^{q}\prod\limits_{p}{(1-\chi _{k,i,l}^{p})}s_{k,i,l}^{{}}}}\le C{{P}_{q,cloud}}\times TD,\\&\quad\quad\forall q\in \mathbb{Q},p\in \mathbb{Q}\backslash \{q\} \\ 
& (c5)\tau _{k,i,l}^{q}+o_{k,i,l}^{q}\le 1,o_{k,i,l}^{q}\in \{0,1\},\forall q\in \mathbb{Q} \\ 
& (c6)\sum\limits_{k\in {{\Re }_{q}}}{\sum\limits_{l=1}^{L}{o_{k,i,l}^{q}p_{k,i,l}^{q}\le {{C}_{q}}}}\times TD,\forall q\in \mathbb{Q} \\ 
\end{aligned}
\end{equation}
The optimized problem is a 0-1 integer programming problem for the variables $\tau _{k,i,l}^{q}$ and $o_{k,i,l}^{q}$. To reduce the computational complexity, equation (21) can be simplified as follows.

\begin{figure}[!h]
	\centering
	\includegraphics[width=9cm]{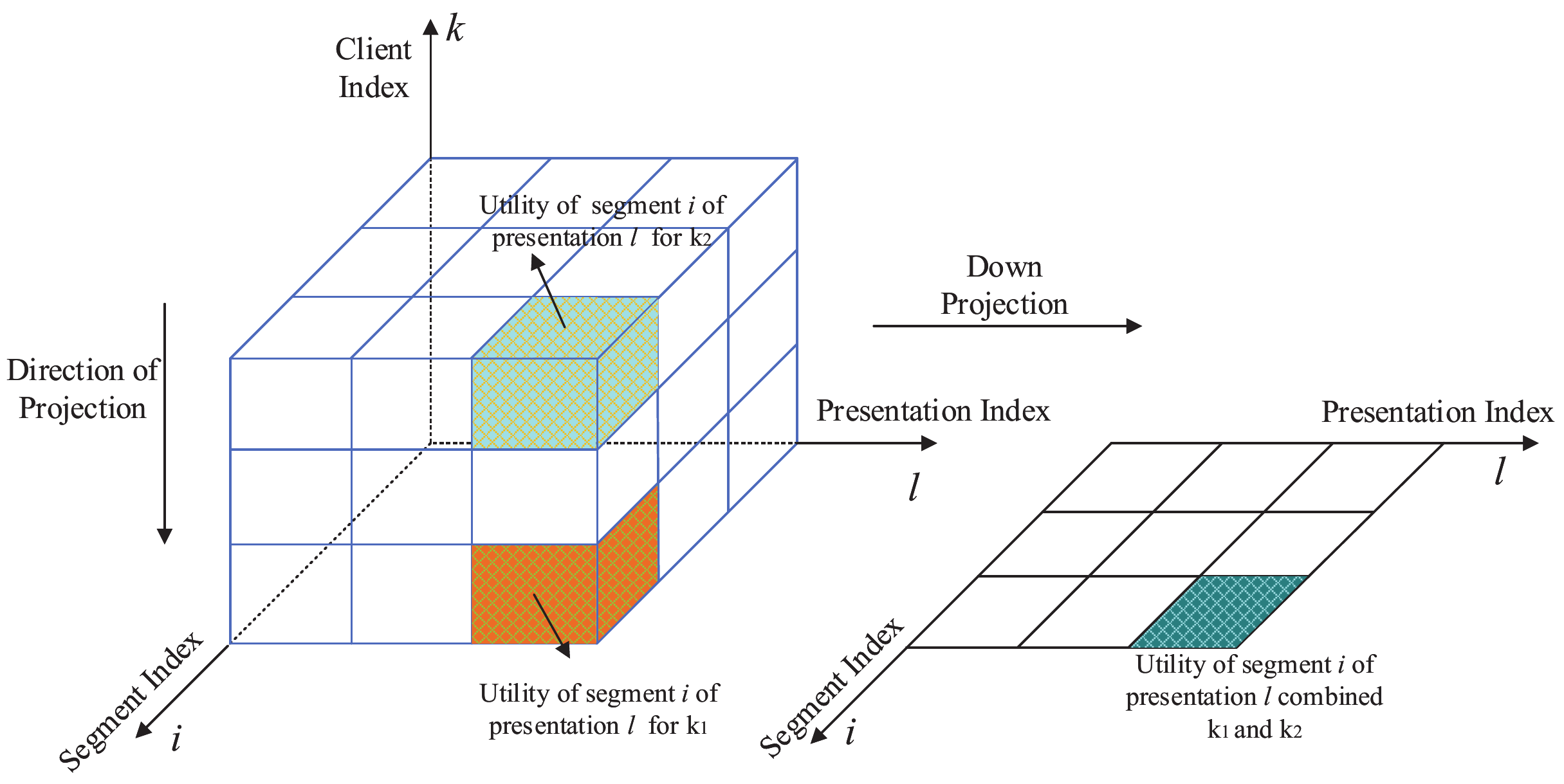}
	\caption{Narrow the Search Range}
	\label{fig:narrow}
\end{figure}

First, we can utilize constraint (c2) to narrow the search range. The client set ${{E}_{q}}$ satisfying $\chi _{k,i,l}^{q}=1,\forall k\in {{\Re }_{q}},1\le l\le L$ can be removed directly from the client set ${{\Re }_{q}}$. In cache decision determination, the condition where one segment is requested by many clients often occurs. To further narrow the search range, we transform the client’s cache and transcoding problem into the segment cache and transcoding problem illustrated in Fig. \ref{fig:narrow}.

When the segment is requested by several clients, the utility function is derived as follows.
\begin{equation}
NU(\tau _{i,l}^{q},o_{i,l}^{q})=\sum\limits_{k\in {{\Re }_{q}}-{{E}_{q}}}^{{}}{U(\tau _{k,i,l}^{q},o_{k,i,l}^{q},{\bm{a}_{k}},\bm{b}_{k}^{*})}
\end{equation} 
 
Then, the search range can be narrowed from $\sum\limits_{q\in \mathbb{Q}}{({{\Re }_{q}}-{{E}_{q}})\times L}$ to $\sum\limits_{q\in \mathbb{Q}}{{{I}_{q}}\times L}$, where ${{I}_{q}}$ is the number of requested segments. Furthermore, the objective function can be simplified as
\begin{equation}
  \max \sum\limits_{q\in \mathbb{Q}}{\sum\limits_{i=1}^{{I}_q}{\sum\limits_{l=1}^{L}{NU(\tau _{i,l}^{q},o_{i,l}^{q})}}}
\end{equation} 
                      
Since $\chi _{k,i,l}^{p}s_{k,i,l}^{{}}$ and $\prod\limits_{p}{(1-\chi _{k,i,l}^{p})}{{s}_{k,i,l}}$ in constraints (c3) (c4) are known constants in the update period, we employ constants $A_{i,l}^{p}$ and ${{B}_{i,l}}$ to substitute into equation (21), which can be simplified as
\begin{equation}
\begin{aligned}
& \underset{\bm{\tau} ,\bm{o}}{\mathop{\max }}\,\sum\limits_{q\in \mathbb{ Q}}{\sum\limits_{i=1}^{{{I}_{q}}}{\sum\limits_{l=1}^{L}{NU(\tau _{i,l}^{q},o_{i,l}^{q})}}} \\ 
& \text{s}\text{.t}\text{.} \\ 
& (c1)\sum\limits_{i=1}^{{{I}_{q}}}{\sum\limits_{l=1}^{L}{\tau _{i,l}^{q}s_{i,l}^{{}}\le S{{C}_{q}}}},\tau _{i,l}^{q}\in \{0,1\},\forall q\in \mathbb{Q} \\ 
& (c2)\sum\limits_{i=1}^{{{I}_{q}}}{\sum\limits_{l=1}^{L}{\tau _{i,l}^{q}A_{i,l}^{p}}}\le C{{P}_{q,p}}\times TD,\forall q\in \mathbb{Q},p\in \mathbb{Q}\backslash \{q\} \\ 
& (c3)\sum\limits_{i=1}^{{{I}_{q}}}{\sum\limits_{l=1}^{L}{\tau _{i,l}^{q}{{B}_{i,l}}}}\le C{{P}_{q,cloud}}\times TD,\forall q\in \mathbb{Q},\text{ }p\in \mathbb{Q}\backslash \{q\} \\ 
& (c4)\tau _{i,l}^{q}+o_{i,l}^{q}\le 1,o_{i,l}^{q}\in \{0,1\},\forall q\in \mathbb{Q} \\ 
& (c5)\sum\limits_{i=1}^{{{I}_{q}}}{\sum\limits_{l=1}^{L}{o_{i,l}^{q}p_{i,l}^{q}\le {{C}_{q}}}}\times TD,\forall q\in \mathbb{Q} \\ 
\end{aligned}
\end{equation}
To solve the problem in equation (24), we employ the brand and branch method \cite{Norkin} to obtain the optimal solution. The detailed process of our proposed algorithm is described in Algorithm 1.
\begin{algorithm}[!h]
	\caption{Solution to the optimal cache and transcoding strategy}
	\label{alg1}
	\textbf{Initialize:} the cache variable $\tau _{i,l}^{q}=0$ and transcoding variable $o_{i,l}^{q}=0$, for $1\le i\le I_q, 1\le l\le L, q\in \mathbb{Q}$. Initialize the size of MEC cache $S{{C}_{q}}$, the size of segment ${{s}_{i,l}}$, the stored segment on the $p$ MEC server $\chi _{i,l}^{p}$ , the transmission bandwidth between the MEC servers $q$ and $p$, the transmission bandwidth between the MEC server $q$ and the cloud server $C{{P}_{q,cloud}}$, the updating time duration $TD$ and the calculation capacity of the $q$ MEC server ${{C}_{q}}$, for $1\le i\le I_q,1\le l\le L,q\in \mathbb{Q}$, $p\in \mathbb{Q}\backslash \{q\}$. 
	
	Let denote the 0-1 integer programming problem (24) by PA, and the corresponding linear programming problem is PB. 

	\textbf{Output:} $\tau _{i,l}^{q}$, $o_{i,l}^{q}$, for $1\le i\le I_q, 1\le l\le L, q\in \mathbb{Q}$.
	
	Transform $\tau _{i,l}^{q}$ and $o_{i,l}^{q}$to the slack variable $0\le \tau _{i,l}^{q}\le 1,0\le o_{i,l}^{q}\le 1$, for $1\le i\le I_q, 1\le l\le L$;
	
	Solve the PB and obtain the optimal solution ${{X}^{*}}$ based on Simplex Algorithm;
	
	Select a feasible solution from PA and use its objective function value as the lower bound of ${{X}^{*}}$;
	
	\While{elements in ${{X}^{*}}$ don't satisfy $(\tau _{i,l}^{q},o_{i,l}^{q})=(0,0)$, $(0,1)$ or $(1,0)$, $1\le i\le I_q, 1\le l\le L$}
	{
		Select one $(\tau _{i,l}^{q},o_{i,l}^{q})$in PB that does not satisfy the 0-1 integer condition, and add the constraints $(\tau _{i,l}^{q},o_{i,l}^{q})=(0,0)$, $(\tau_{i,l}^{q},o_{i,l}^{q})=(0,1)$and $(\tau _{i,l}^{q},o_{i,l}^{q})=(1,0)$to the PB to formulate the subproblem PB$_1$, PB$_2$ and PB$_3$;
		
		Employ the Simplex Algorithm to solve the PB$_1$, PB$_2$ and PB$_3$ without considering the 0-1 integer constraints;
		
		Select the maximum objective function value from the solutions of PB$_1$, PB$_2$ and PB$_3$ as the new upper bound of ${{X}^{*}}$; 
		
		Select the minimum objective function value which satisfies the 0-1 integer condition from each subproblem as the new lower bound of ${{X}^{*}}$;
		
		Delete the subproblem whose objective function value does not belong to the interval between the lower bound and upper bound; 
		
		Update the optimal solution ${{X}^{*}}$;
		
	}
\end{algorithm}

\subsubsection{$a_{k}^{*}$ Determination with Known $(\bm{\tau} _{k}^{*},\bm{o}_{k}^{*},\bm{b}_{k}^{*})$}
The derivation of $\bm{a}_{k}^{*}$ in the previous section is conducted to determine the optimal $(\bm{\tau} _{k}^{*}, \bm{o}_{k}^{*}, \bm{b}_{k}^{*})$ based on the fixed RB set $N_{k}^{{{h}_{q}}}$. The goal is to calculate the utility improvement $CI_{k,n}^{{{h}_{q}}}$. In this section, we determine the assignment strategy of RB $n$ according to the value of $CI_{k,n}^{{{h}_{q}}}$. To achieve $\max \sum_{q}{\sum_{k}{\sum_{l}{U({\bm{\tau }_{k}},{\bm{o}_{k}},{\bm{a}_{k}},{\bm{b}_{k}})}}}$, we need to assign RB $n$ to the client with the maximum $CI_{k,n}^{{{h}_{q}}}$ as follows:
\begin{equation}
a{{_{k,n}^{{{h}_{q}}}}^{*}}=\left\{ \begin{aligned}
& 1,k=\underset{k\in \mathbb{Z}({{h}_{q}})}{\mathop{\arg \max }}\,CI_{k,n}^{{{h}_{q}}} \\ 
& 0,otherwise \\ 
\end{aligned} \right.
\end{equation}

Based on the above analysis, the criteria used to determine the RB assignment, MCS selection, video segment cache and transcoding have now been obtained. The optimization process is executed in every updated period, and our solution procedures are summarized in Algorithm 2 named CTRA (Cache, Transcoding and Resource Allocation).
\begin{algorithm}[!h]
	\caption{CTRA}
	\label{alg2}
	\textbf{Initialize:} the total available RB set $\Omega ({{h}_{q}})=\{1,...,{{N}_{{{h}_{q}}}}\}$ and the set of the RBs assigned to client k, $N_{k}^{{{h}_{q}}}=\varnothing$, for $k\in \mathbb{Z}({{h}_{q}}), {{h}_{q}}\in {{\mathcal{H}}_{q}}, q\in \mathbb{Q}$. Initialize the RB assignment indicator and the MCS indicator $a_{k,n}^{{{h}_{q}}}=0$, $b_{k,m}^{{{h}_{q}}}=0$, for $k\in \mathbb{Z}({{h}_{q}}), n\in \Omega ({{h}_{q}}), {{h}_{q}}\in {{\mathcal{H}}_{q}}, q\in \mathbb{Q}$.
	
	\textbf{Output:} $\tau _{i,l}^{q}$, $o_{i,l}^{q}$, $a_{k,n}^{{{h}_{q}}}$, $b_{k,m}^{{{h}_{q}}}$, for $1\le i\le I_q, 1\le l\le L, k\in \mathbb{Z}({{h}_{q}}), {{h}_{q}}\in {{\mathcal{H}}_{q}}, q\in \mathbb{Q}$.
	
	\For{$q\in \mathbb{Q}$}
	{
		\For{${{h}_{q}}\in {{\mathcal{H}}_{q}}$}
		{
		\While{there are RBs unassigned and packets left in MAC queue ($\Omega ({{h}_{q}})\ne \varnothing $ and $\mathbb{Z}({{h}_{q}})\ne \varnothing $)}
			{
			\For{$n\in \Omega ({{h}_{q}})$}
				{
					\For{$k\in \mathbb{Z}({{h}_{q}})$}
						{
							Update the RB assignment set, $N_{k}^{{{h}_{q}}}=N_{k}^{{{h}_{q}}}\bigcup \{n\}$;
							
							Obtain optimal MCS selection $b{{_{k,m}^{{{h}_{q}}}}^{*}}$with fixed $N_{k}^{{{h}_{q}}}$;
							
							Obtain the optimal cache variable and transcoding variable, $\tau _{i,l}^{q}$,$o_{i,l}^{q}$ based on Algorithm 1;
							
							Calculate the utility level $U({\bm{\tau }_{k}}^{*},{\bm{o}_{k}}^{*},{\bm{a}_{k}},{\bm{b}_{k}}^{*})$ based on Eq.(17);
							
							Calculate the value of the cost function $C{{I}_{k,n}}$ based on Eq.(19);	
						}
					Assign RB $n$ to the client with the maximum $C{{I}_{k,n}}$, ${{k}^{*}}=\underset{k\in \mathbb{Z}({{h}_{q}})}{\arg \max }\,C{{I}_{k,n}}$;
					
					Update the RB assignment set $N_{k}^{{{h}_{q}}}=N_{k}^{{{h}_{q}}}\backslash \{n\}$, RB assignment indicator $a{{_{k,n}^{{{h}_{q}}}}^{*}}=0$ and MCS index $b{{_{k,m}^{{{h}_{q}}}}^{*}}=0$ for $k\ne {{k}^{*}}$; 
					
					Remove RB $n$ from the available RB set, $\Omega ({{h}_{q}})=\Omega ({{h}_{q}})\backslash \{n\}$; 
					
					\If{all the packets of client $k$ can be transmitted, $\sum\limits_{n\in N_{{{k}^{*}}}^{{{h}_{q}}}}{a{{_{k,n}^{{{h}_{q}}}}^{*}}r_{m}^{*}\ge L{{S}_{{{k}^{*}}}}}$}
						{
							Remove client ${{k}^{*}}$ from the set $\mathbb{Z}({{h}_{q}})$.	
						}
				}

			}

		}

	}
\end{algorithm}

\section{Simulation Results}
\subsection{Parameter Setting}
In this section, we modified the Vienna LTE-A simulation platform by adding cloud server, MEC server and joint optimization algorithm of the video segment cache, transcoding and wireless resource allocation. The detailed parameter settings are presented in Table \ref{tab:parameter}.
\begin{table}[!h]\scriptsize
	\caption{Simulation Parameter Configuration}\label{tab:parameter}
	\centering
	\begin{tabular}{|l|l|}
		\hline
		\multirow{2}{*}{Position of eNodeBs}                                                                & \multirow{2}{*}{\begin{tabular}[c]{@{}l@{}}{[}600, 342{]},{[}600, -342{]}, {[}0, -690{]}, \\ {[}-600, -342{]}, {[}-600 342{]}, {[}0, 690{]} (m)\end{tabular}} \\
		&                                                                                                                                                               \\ \hline
		Bandwidth                                                                                           & 20MHz                                                                                                    \\ \hline
		Transmitting power of eNodeB                                                         & 100 dBm                                                                                                                                                  \\ \hline
		Position of MEC servers                                                                             & {[}-600,0{]}, {[}0,0{]}, {[}600, 0{]} (m)                                                                                                                     \\ \hline
		\begin{tabular}[c]{@{}l@{}}Transmission capacity between\\ MEC server and cloud server\end{tabular} & 500Mbps                                                                                             \\ \hline

		\begin{tabular}[c]{@{}l@{}}Transmission capacity among \\ MEC servers\end{tabular}                  & 200Mbps                                                                                                                                                       \\ \hline
			Computing capacity of MEC server                                                        & 3.6 G Cycles                                                                                                                                                 \\ \hline
			Channel model                                                      & Typical urban                                                                                                                                                 \\ \hline
			Propagation model                                                      & Marco-cell urban                                                                                                                                                 \\ \hline
		Pathloss                                                                                            & 20dB                                                                                                                                                          \\ \hline
		The number of clients                                                                                 & 378                                                                                                                                                           \\ \hline
		TTI (Transport Time Interval)                                                                                                 & 1 ms                                                                                                                                                           \\ \hline
		Simulation time                                                                                                 & 60000 ms                                                                                                                                                           \\ \hline
		Video segment encoder                                                                               & H.264/AVC                                                                                                                                                     \\ \hline
		\multirow{2}{*}{Video sequence}                                                                     & \begin{tabular}[c]{@{}l@{}}CIF format: Bus,Coastguard,Highway,\\ Flower, Foreman, Crew, News, Soccer\end{tabular}                                             \\ \cline{2-2} 
		& \begin{tabular}[c]{@{}l@{}}720P format: ParkingLot1,Stockholm,\\ Mobcal, In\_to\_tree, Park\_joy, Shields, \\ Ducks\_take\_off,Old\_town\_cross\end{tabular} \\ \hline
		\begin{tabular}[c]{@{}l@{}}The number of frames \\ in one segment\end{tabular}                      & 60                                                                                                                                                            \\ \hline
		Frame rate                                                                                              & 30 fps                                                                                                                                                           \\ \hline
		Width$\times$Height                                                                                 & 352$\times$288,1280$\times$720                                                                                                                                \\ \hline
		QP                                                                                                  & 28,29,30,31,32                                                                                                                                               \\ \hline
		version                                                                                      & 1,2,3,4,5                                                                                                                                                     \\ \hline
		Experimental parameters                                                                             & \begin{tabular}[c]{@{}l@{}}$\alpha=0.5$, $\zeta=0.8$, $TD=50$ ms, \\ $\omega=2$, $\mu=30$ Cycle/byte, $\omega =2$, \\
		${{\theta }_{1}}=0.125$, ${{\theta }_{2}}=1$, ${{\theta }_{3}}=1$, \\ ${{\theta }_{4}}=0.025$
		\end{tabular}                                        \\ \hline
	\end{tabular}
\end{table}

Since the number of test videos and the CPU processing speed are limited, we reduce the cache capacity and computing capacity of the MEC server and the number of requested videos compared with the actual scenario. The test videos are downloaded from the websites of \cite{YUV} and \cite{Xiph}. The test video with the lowest video rate is Highway, while the video with the highest rate is Park$\_$joy. The initial video popularity follows a Zipf distribution with parameter $\beta =0.6$, whereas the video request arrival rate follows a Poisson distribution with rate $\lambda =0.8$. The algorithms compared in this section are as follows:

(1)	CTRA: The proposed joint optimization algorithm in this paper.

(2)	LRU-RR (Least Recently Used and Round Robin) \cite{Jacob}\cite{Al}: The cache strategy and the transcoding strategy are based on the video segment popularity in the previous period. The resource allocation is based on RR algorithm.

(3)	LFU (Least Frequently Used and Round Robin) \cite{Jacob}\cite{Al}: The cache strategy and the transcoding strategy are based on the video segment popularity up to the current period. The resource allocation is based on RR algorithm.

(4)	RBCC (Retention-based Collaborative Caching) \cite{Mehrabi_QoE}: It takes into account the values of the requested video segment from the neighborhood clients for caching at each local edge server to improve client's QoE without considering the transcoding. The resource allocation is based on the request video segment rate. 

(5)	Greedy$\_$MSMC (Greedy$\_$MEC Servers Map Clients) \cite{Mehrabi}: The goal of this work is to maximize the QoE of total clients and proportional fairness of the bitrate allocation without considering the transcoding. 

\subsubsection{Performance Evaluation}
We validate the performance in terms of client's  throughput, system backhaul traffic, video quality, presentation switch and hit ratio of video segment and the playback rebuffering time under different simulation configurations. 

\begin{figure}[!h]
	\centering
	\includegraphics[width=9cm]{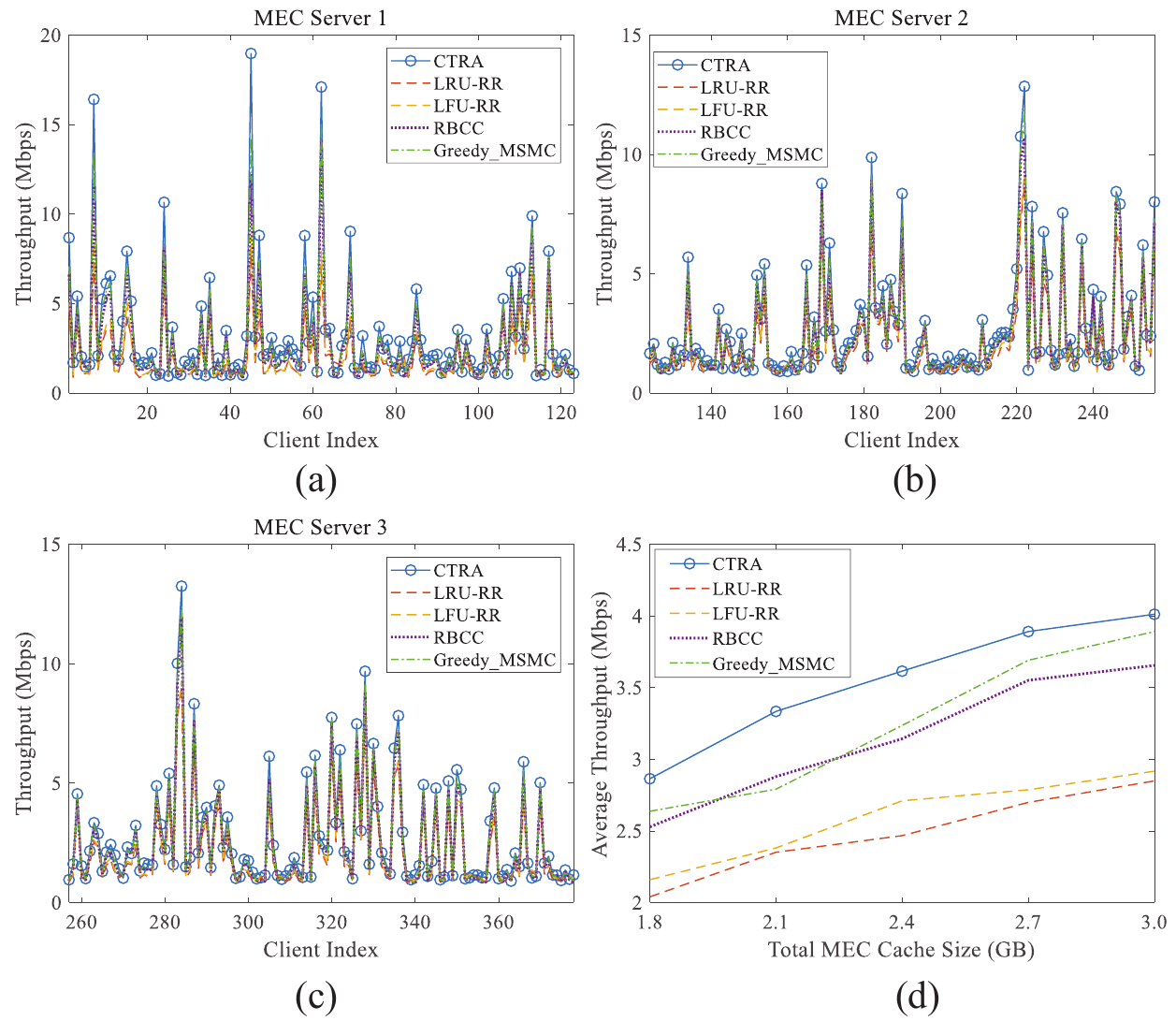}
	\caption{(a)-(c) Throughput of each client for different update algorithms served by MEC servers 1, 2 and 3. (d) Average throughput of all clients for different update algorithms vs. different total MEC storage sizes.}
	\label{fig:throughput}
\end{figure}

In Fig. 5. (a), (b) and (c), we compare client throughput when the cache space of MEC servers 1, 2 and 3 is 650 MB, 600 MB and 550 MB, respectively. Compared with that of the other four algorithms, the clients' throughput for the CTRA algorithm is the highest overall because it comprehensively considers the sizes of the requested segments and the client's current remaining buffer time and dynamically adjusts the segment update strategy and resource allocation strategy to ensure that the client's throughput can meet the smooth playback requirements. Fig. 5. (d) presents the average throughput for different total MEC cache sizes. In the initial stage, the average throughput of the CTRA algorithm is the highest, followed by that of the Greedy$\_$MSMC and RBCC algorithms and finally the LRU-RR algorithm. As the MEC cache space increases, the average throughput of the Greedy$\_$MSMC and RBCC algorithms is relatively close, alternately leading in different MEC cache sizes, but always outperforming the LFU-RR and LRU-RR algorithms. When the total MEC cache space is 3.0 GB, the average throughput of each algorithm reaches the maximum, and the average throughput of the CTRA algorithm increases by 40.82\%, 37.54\%, 9.77\% and 3.11\% compared with that of LRU-RR, LFU-RR, RBCC and Greedy$\_$MSMC, respectively. The results show that as the MEC cache space increases, more different presentations of segments can be cached to meet the requirements of different clients to improve their throughput. In addition, the CTRA algorithm can adjust the video segment cache, transcoding and wireless resource allocation dynamically in different MEC cache spaces, which can effectively improve the client's throughput.

\begin{figure}[!h]
	\centering
	\includegraphics[width=9cm]{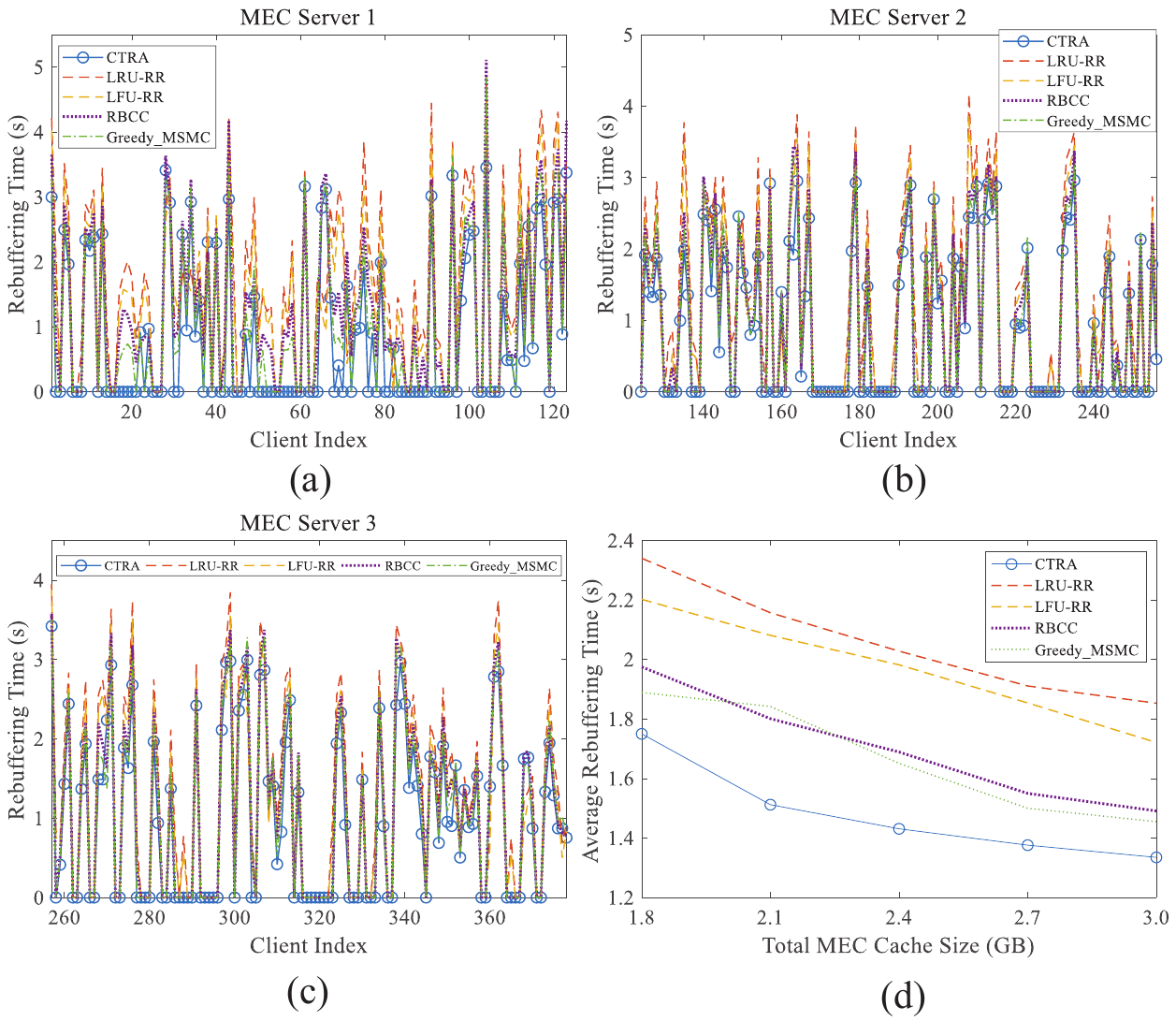}
	\caption{(a)-(c) Rebuffering time of each client for different update algorithms served by MEC servers 1, 2 and 3. (d) Average rebuffering time of all clients for different update algorithms vs. different total MEC cache sizes.}
	\label{fig:rebuffer}
\end{figure}

In Fig. 6. (a), (b) and (c), we compare the client’s rebuffering time when the cache space of MEC servers 1, 2 and 3 is 650 MB, 600 MB and 550 MB, respectively. Compared with the other four algorithms, the clients’ rebuffering time with the CTRA algorithm is the highest overall because it comprehensively considers the client’s current channel state and remaining buffer time to cache or transcode the approximate presentation of video segments to ensure smooth playback. Fig. 6. (d) presents the average rebuffering time for different total MEC cache sizes. In the initial stage, the average rebuffering time of the CTRA algorithm is the lowest, followed that of the Greedy$\_$MSMC and RBCC algorithms and finally the LRU-RR algorithm. As the MEC cache space increases, the average rebuffering times of the Greedy$\_$MSMC and RBCC algorithms are relatively close, alternately leading in different MEC cache sizes, but always lower than that of the LFU-RR and LRU-RR algorithms. When the total MEC cache space is 3.0 GB, the average rebuffering time of each algorithm reaches the minimum, and the average rebuffering time of the CTRA algorithm is reduced by 27.95\%, 22.43\%, 10.46\% and 8.25\% compared with LRU-RR, LFU-RR, RBCC and Greedy$\_$MSMC, respectively. The results show that as the MEC cache space increases, the CTRA algorithm can achieve a shorter rebuffering time by means of the effective video segment cache, transcoding and wireless resource allocation.

\begin{figure}[!h]
	\centering
	\includegraphics[width=9cm]{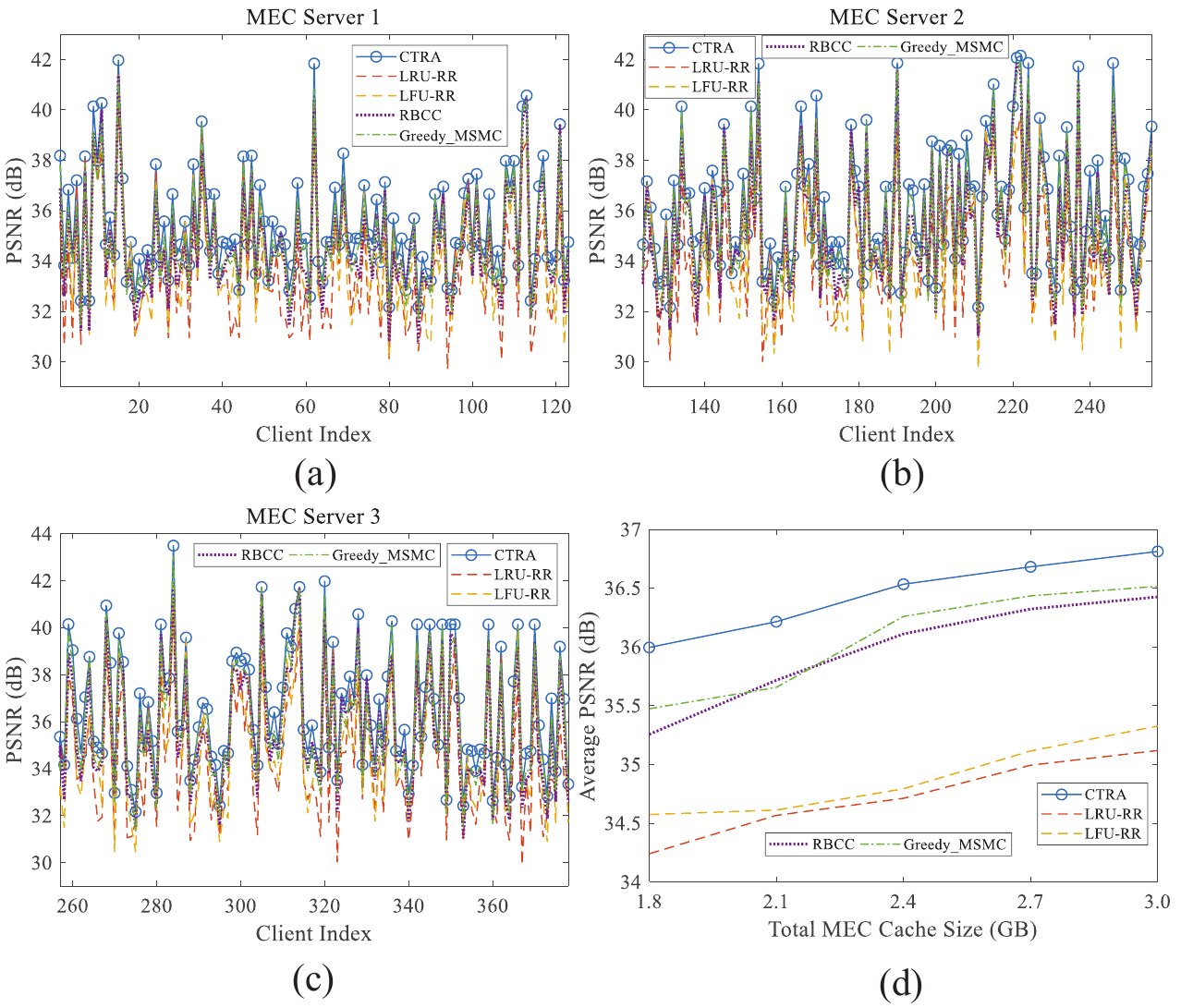}
	\caption{(a)-(c) PSNR of received video segments of each client for different update algorithms served by MEC servers 1, 2 and 3. (d) Average PSNR of received video segments of all clients for different update algorithms vs. different total MEC cache sizes.}
	\label{fig:PSNR}
\end{figure}

In Fig. 7. (a), (b) and (c), we compare the PSNR of received video segments of each client when the cache of MEC servers 1, 2 and 3 is 650 MB, 600 MB and 550 MB, respectively. Compared with the other algorithms, although some clients' received video quality with the CTRA algorithm is lower, the received video quality of most clients is higher. Since the utility function model of the CTRA algorithm considers the effect of the PSNR of received video frames on the client's QoE, the video quality of most clients can be maintained at a high level. Fig. 7. (d) presents the average PSNR of all clients with different total MEC cache sizes. As the MEC cache space increases, the average PSNR of all clients for each algorithm grows gradually, and the CTRA algorithm always maintains the highest PSNR.

\begin{figure}[!h]
	\centering
	\includegraphics[width=9cm]{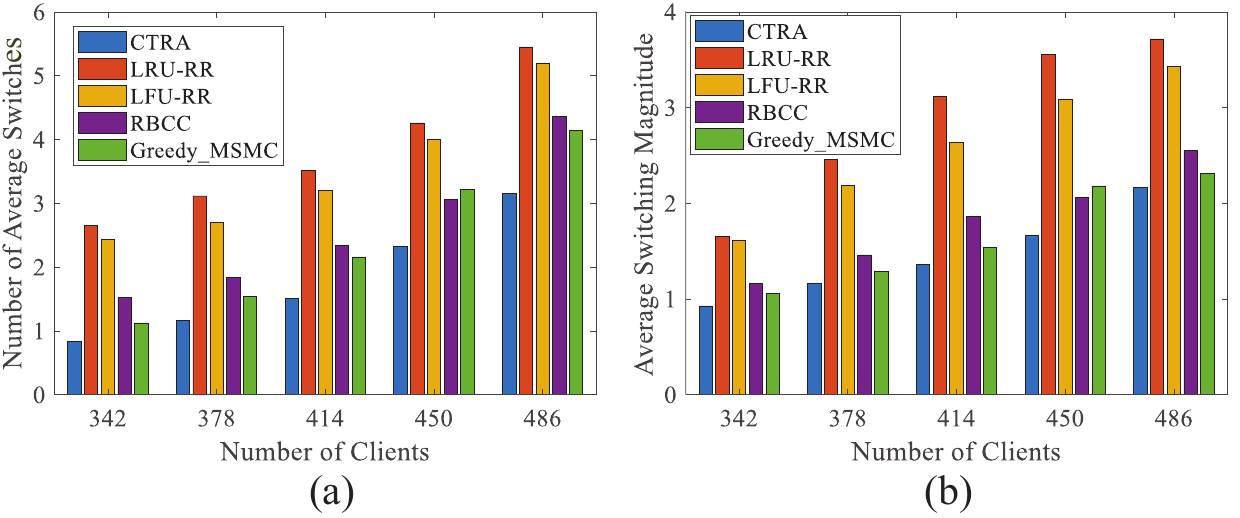}
	\caption{(a) Average number of switches vs. different number of clients. (b) Average switching magnitude vs. different number of clients.}
	\label{fig:switch}
\end{figure}

Fig. 8. (a)(b) present the number of average switches and average switching magnitudes for video segment presentation with different numbers of clients when the total MEC cache space size is 1.8 GB. As the number of clients increases, the average number of switches and average switching magnitude of video segments of each algorithm increase gradually because more clients compete for limited cache, transcoding and wireless resources. The MEC server has difficulty consistently providing the same presentation of video segments for clients, so clients must receive different presentations of video segments to ensure smooth playback. When the number of clients is higher than 414, the average switching times and switching magnitude of the RBCC and Greedy$\_$MSMC algorithms alternately lead but are always lower than those of the LRU-RR and LFU-RR algorithms. This is because the RBCC and Greedy$\_$MSMC algorithms take into account the client’s experience and reduce the video quality fluctuation caused by the presentation switch.

\begin{figure}[!h]
	\centering
	\includegraphics[width=9cm]{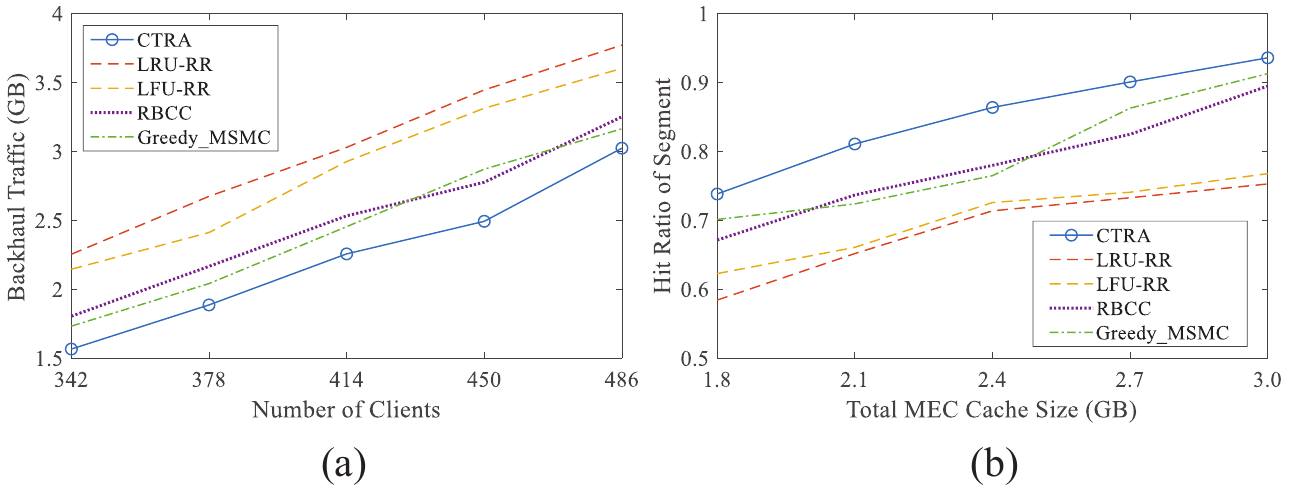}
	\caption{(a) Backhaul traffic of MEC servers for different update algorithms vs. different numbers of clients. (b) Hit ratio of segment of all clients for different update algorithms vs. different total MEC cache sizes.}
	\label{fig:back}
\end{figure}

We present the backhaul traffic of MEC servers with different numbers of clients and hit ratios of segments with different total MEC cache sizes in Fig. 9. In Fig. 9. (a), when the total MEC cache space is 1.8 GB, due to the limited MEC cache space and the increase in requested video segments, the MEC server needs to cache and delete video segments frequently, which results in an increase in backhaul traffic between the MEC server and the cloud server. Fig. 9. (b) shows that as the MEC cache size increases, the segment hit ratio of each algorithm gradually increases because the increase in the MEC cache size ensures that more presentations of the segment can be precached in the MEC server. When the client requests a new segment, the eNodeB can directly download the required segment from the MEC server and deliver it to the client. The hit ratio of the video segment of the CTRA algorithm remains the highest throughout the entire process because the cooperation mechanism can help the local MEC server download the requested segments from the neighboring MEC servers quickly.

\section{Conclusion}
In this paper, we propose a joint optimization of MEC-assisted video segment cache, transcoding and wireless resource allocation. First, we predeploy the MEC cache based on historical video request data to improve the utilization of the MEC cache space. The popularity of video segments based on the number of requests and the client's channel state is employed to divide the MEC cache area and obtain the video segment delete strategy. Then, the client's channel state, playback status and cooperation mechanism among MEC servers are combined to calculate the client's priority, presentation switching magnitude, continuous playback time and PSNR of received video frames. Based on the above four factors, a utility function model based on the client’s QoE is established. In each update period, with the constraints of MEC storage capacity, transcoding capacity, transmission bandwidth among MEC servers and the transmission capacity of the eNodeB, the joint optimization of video segment cache, transcoding and wireless resource allocation is formulated as a mixed-integer nonlinear programming mathematical model to maximize the total utility of the clients. Finally, a low-complexity heuristic algorithm is proposed to decompose the original problem into multiple subproblems. We provide numerical simulation results in terms of system throughput, system backhaul traffic, video quality, hit ratio of video segments and playback rebuffering time. The results show that our proposed algorithm outperforms the LRU, LFU, WGDSF and RBCC algorithms.

\section*{Acknowledgment}
This research work was supported in part by the National Science Foundation of China (61701389
, U1903213) and the Shaanxi Key R\&D Program (2018ZDCXL-GY-04-03-02).


\end{document}